\documentclass[12pt]{article}
\usepackage{amsmath}
\usepackage{graphics}
\usepackage{graphicx,psfrag,epsf}
\usepackage{enumerate}
\usepackage{natbib}
\usepackage{url}

\let\proglang=\textsf
\let\code=\texttt

\begin{document}

\bibliographystyle{natbib}

\def\spacingset#1{\renewcommand{\baselinestretch}
{#1}\small\normalsize} \spacingset{1}


\title{\bf Computing and Graphing Probability Values of Pearson Distributions: A \proglang{SAS/IML} Macro}
	\author{Wei Pan\\Duke University\\\\Xinming An\\SAS Institute Inc.\\\\Qing Yang\\Duke University\\\date{April 2017}}
\maketitle

\bigskip
\begin{abstract}
Any empirical data can be approximated to one of Pearson distributions using the first four moments of the data \citep{Elderton69, Pearson95, Solomon78}. Thus, Pearson distributions made statistical analysis possible for data with unknown distributions. There are both extant old-fashioned in-print tables \citep{Pearson72} and contemporary computer programs \citep{Amos71, Bouver74, Bowman79, Davis83, Pan09} available for obtaining percentage points of Pearson distributions corresponding to certain \textit{pre-specified} percentages (or probability values) (e.g., 1.0\%, 2.5\%, 5.0\%, etc.), but they are little useful in statistical analysis because we have to rely on unwieldy second difference interpolation to calculate a probability value of a Pearson distribution corresponding to \textit{any} given percentage point, such as an observed test statistic in hypothesis testing. Thus, the present study develops a \proglang{SAS/IML} macro program to compute and graph probability values of Pearson distributions for any given percentage point so as to facilitate researchers to conduct statistical analysis on data with unknown distributions.
\end{abstract}

\noindent%
\vfill

\newpage
\spacingset{1}
\section{Introduction}
\label{sec:intro}

Most of statistical analysis relies on normal distributions, but this assumption is often difficult to meet in reality. Pearson distributions can be approximated for any data using the first four moments of the data \citep{Elderton69, Pearson95, Solomon78}. Thus, Pearson distributions made statistical analysis possible for any data with unknown distributions. For instance, in hypothesis testing, a sampling distribution of an observed test statistic is usually unknown but the sampling distribution can be fitted into one of Pearson distributions. Then, we can compute and use a \textit{p}-value (or probability value) of the approximated Pearson distribution to make a statistical decision for such distribution-free hypothesis testing.

There are both extant old-fashioned in-print tables \citep{Pearson72} and contemporary computer programs \citep{Amos71, Bouver74, Bowman79, Davis83, Pan09} that provided a means of obtaining percentage points of Pearson distributions corresponding to certain \textit{pre-specified} percentages (or probability values) (e.g., 1.0\%, 2.5\%, 5.0\%, etc.). Unfortunately, they are little useful in statistical analysis because we have to employ unwieldy second difference interpolation for both skewness $\surd\beta_{1}$ and kurtosis $\beta_{2}$ to calculate a probability value of a Pearson distribution corresponding to \textit{any} given percentage point, such as an observed test statistic in hypothesis testing. Thus, a new program is needed for easily computing probability values of Pearson distributions for any given probability values; and therefore, researchers can utilize the program to conduct more applicable statistical analysis, such as distribution-free hypothesis testing, on data with unknown distributions.

\section{Pearson distributions}
\label{sec:distr}

Pearson distributions are a family of distributions which consist of seven different types of distributions plus normal distribution (Table \ref{table1}). Let $X$ represent given data, once its first four moments are calculated by 

\begin{equation}
\left\{
\begin{array}{l}
\mu_{1}'=E(X); \\
\mu_{i}=E[X-E(X)]^{i}=E[X-\mu_{1}']^{i}, i=2,3,4,
\end{array}
\right.\label{1}
\end{equation}

\noindent%
types of Pearson distributions to which $X$ will be approximated can be determined by a $\kappa$-criterion that is defined as follows \citep{Elderton69}:

\begin{equation}
\kappa=\frac{\beta_{1}(\beta_{2}+3)^{2}}{4(4\beta_{2}-3\beta_{1})(2\beta_{2}-3\beta_{1}-6)},
\label{2}
\end{equation}

\noindent%
where the $\beta$-coefficients (\textit{i.e.}, skewness and kurtosis) are calculated as follows:

\begin{equation}
\left\{
\begin{array}{l}
\surd\beta_{1}=\frac{\mu_{3}}{\mu_{2}^{3/2}} \mbox{ (also } \beta_{1}=(\surd\beta_{1})^{2}=\frac{\mu_{3}^{2}}{\mu_{2}^{3}}); \\
\beta_{2}=\frac{\mu_{4}}{\mu_{2}^{2}}.
\end{array}
\right.\label{3}
\end{equation}

\begin{table}[h]
	\centering
	\begin{tabular}{llll}
		\hline
		\multicolumn{1}{c}{Type} & \multicolumn{1}{c}{$\kappa$-Criterion} & \multicolumn{1}{c}{Density Function} & \multicolumn{1}{c}{Domain} \\
		\hline
		\multicolumn{4}{c}{\textit{Main Type}} \\
		\hline
		I      & $\kappa<0$ & $f(x)=y_0(1+\frac{x}{a_1})^{m_1}(1-\frac{x}{a_2})^{m_2}$ & $-a_{1}\leq x \leq a_{2}$ \\ 
		IV     & $0<\kappa<1$ & $f(x)=y_0(1+\frac{x^2}{a^2})^{-m}e^{-\nu\arctan(x/a)}$ & $-\infty< x <\infty$ \\ 
		VI     & $\kappa>1$ & $f(x)=y_0(x-a)^{q_2}x^{-q_1}$ & $a\leq x < \infty$ \\
		\hline
		\multicolumn{4}{c}{\textit{Transition Type}} \\
		\hline
		Normal & $\kappa=0 \mbox{ } (\beta_{2}=3)$ & $f(x)=y_0e^{-x^2/(2\mu_2)}$ & $-\infty< x <\infty$ \\ 
		II     & $\kappa=0 \mbox{ } (\beta_{2}<3)$ & $f(x)=y_0(1-\frac{x^2}{a^2})^m$ & $-a\leq x \leq a$ \\ 
		III    & $\kappa=\pm\infty$ & $f(x)=y_0(1+\frac{x}{a})^{\gamma{a}}e^{-\gamma{x}}$ & $-a\leq x <\infty$ \\ 
		V      & $\kappa=1$ & $f(x)=y_0x^{-p}e^{-\gamma/x}$ & $0< x < \infty$ \\
		VII    & $\kappa=0 \mbox{ } (\beta_{2}>3)$ & $f(x)=y_0(1+\frac{x^2}{a^2})^{-m}$ & $-\infty< x <\infty$ \\
		\hline
	\end{tabular}
	\caption{Types of Pearson distributions.}
	\label{table1}
\end{table}

The determination of types of Pearson distributions by the $\kappa$-criterion (Equation \ref{2}) is illustrated in Table~\ref{table1}. From Table \ref{table1}, we can also see that for each type of Pearson distributions, its density function has a closed form with a clearly defined domain of $X$. The closed form of density functions made numerical integration possible for obtaining probability values of approximated Pearson distributions. Following the calculation formulas introduced in \citet{Elderton69}, the parameters (e.g., $y_{0}$, $m_{1}$, $m_{2}$, $a_{1}$, $a_{2}$, etc.) of the density functions will be automatically computed in a \proglang{SAS/IML} \citep{SAS-IML11} macro program described in the next section. Then, probability values of Pearson distributions can be obtained through numerical integration with the SAS subroutine \proglang{QUAD}. 

\section{A \proglang{SAS/IML} macro program}
\label{sec:macro}

The main \proglang{SAS/IML} macro program to compute and graph probability values of Pearson distributions is as follows:

\begin{center}
	\code{\%PearsonProb(mu2 = , mu3 = , mu4 = , x0 = , plot = )}
\end{center}

\noindent%
where 

\begin{list}{}{}
	\item \code{mu2} = the second moment $\mu_2$;
	\item \code{mu3} = the third moment $\mu_3$;
	\item \code{mu4} = the fourth moment $\mu_4$;
	\item \code{x0} = the percentage point $x_{0}$;
	\item \code{plot} = 1 for graph, 0 for no graph.
\end{list}

It is worth noting that the first moment $\mu_{1}'$ is not an input value for this macro and that only the second, the third, and the fourth moments as well as a percentage point $x_{0}$ and 1 or 0 for  \code{plot} are required. The reason is that the first moment $\mu_{1}'$ has been already included in the calculation for the higher-oder moments (see Equation \ref{1}). 

This \proglang{SAS/IML} macro program starts with computing $\beta$-coefficients defined in Equation \ref{3} using the inputed values of $\mu_2$, $\mu_3$, $\mu_4$, and $x_{0}$. Then, plug them into Equation \ref{2} to calculate $\kappa$. Based on the value of the $\kappa$, a specific type of Pearson distribution is determined by the $\kappa$-criterion displayed in Table \ref{table1}, followed by calculations of the parameters (e.g., $y_{0}$, $m_{1}$, $m_{2}$, $a_{1}$, $a_{2}$, etc.) for the density function of the specific type of Pearson distribution listed in Table \ref{table1}. To compute the probability value of the specific Pearson distribution corresponding to the inputed percentage point $x_{0}$, the SAS subroutine \proglang{QUAD} is called for numerical integration. If the inputed $x_{0}$ is beyond the defined domain, a waring message will be printed as ``\code{WARNING: x0 is out of the domain of type VI Pearson distribution}," for example. Finally, the computed probability value along with the parameters are printed.

\begin{figure}
	\centering
	\includegraphics[width=.8\textwidth]{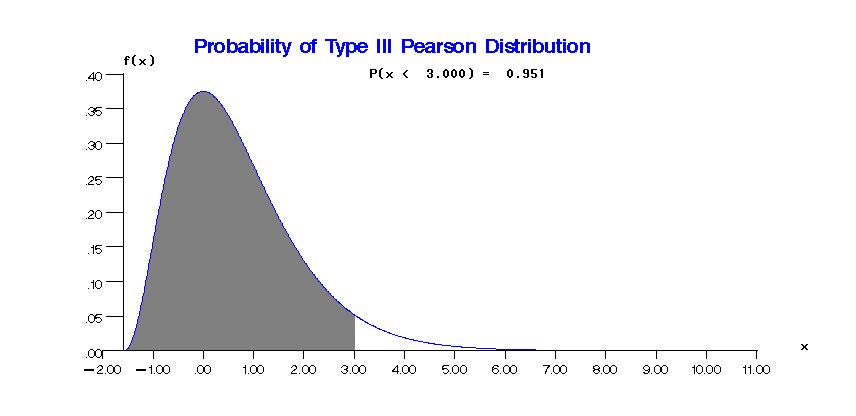}
	\caption{A type III Pearson distribution with a probability value indicated.}
	\label{figure1}
\end{figure}

To graph the probability value on the approximated density fucntion of the Pearson distribution, a small \proglang{SAS/IML} macro \code{\%plotprob} was written for use within the main \proglang{SAS/IML} macro \code{\%PearsonProb(mu2 = , mu3 = , mu4 = , x0 = , plot = )}. If \code{1} is inputed for \code{plot}, the SAS subroutines \proglang{GDRAW}, \proglang{GPLOY}, etc. are called in the small graphing macro for plotting the density function and indicating probability value (Figure \ref{figure1}). Otherwise (i.e., \code{plot = 0}), no graph is produced.  

\section{Evaluation of the program}
\label{sec:eval}

To evaluate the accuracy of the \proglang{SAS/IML} macro program for computing and graphing probability values of Pearson distributions, the calculated parameters of the approximated Pearson distributions from this \proglang{SAS/IML} macro were first compared with the corresponding ones in \citet{Elderton69}. As can be seen in Table \ref{table2}, the absolute differences between the calculated parameters from the \proglang{SAS/IML} macro and those from \citet{Elderton69}'s tables are all very small with almost all of them less than .001 and a few less than .019. The same story applies to the relative differences with an unsurprising exception (4.46\%) of $\kappa$ for type IV whose original magnitude is very small.

\begin{table}[h]
	\centering
	\begin{tabular}{lrrrrr}
		\hline
		&& \multicolumn{1}{c}{Value from} & \multicolumn{1}{c}{Value from Elderton} & \multicolumn{1}{c}{Absolute} & \multicolumn{1}{c}{Relative} \\
		Type$^a$ & Parameter & \multicolumn{1}{c}{\proglang{SAS/IML} Macro} & \multicolumn{1}{c}{and Johnson (1969)} & \multicolumn{1}{c}{Difference$^b$} & \multicolumn{1}{c}{Difference$^c$} \\
		\hline
		I  & $\beta_1$  & .507296    & .507296    & $< .0001$ & $< .01$\% \\ 
		& $\beta_2$  & 2.935111   & 2.935110   & $< .0001$ & $< .01$\% \\ 
		& $\kappa$        & -.264690   & -.264500   & .0002   & .07\% \\ 
		& $r$        & 5.186821   & 5.186811   & $< .0001$ & $<.01$\% \\  
		& $\alpha_1$ & 1.977543   & 1.996380   & .0188   & .94\% \\ 
		& $\alpha_2$ & 13.508428  & 13.527280  & .0189   & .14\% \\ 
		& $m_1$      & .406954    & .409833    & .0029   & .70\% \\ 
		& $m_1$      & 2.779867   & 2.776878   & .0030   & .12\% \\ 
		\hline
		IV & $\beta_1$  & .005366    & .005366    & $< .0001$ & $< .01$\% \\ 
		& $\beta_2$  & 3.172912   & 3.172912   & $< .0001$ & $< .01$\% \\ 
		& $\kappa$        & .012230    & .012800    & .0006   & 4.46\% \\ 
		& $r$        & 39.442562  & 39.442540  & $< .0001$ & $< .01$\% \\  
		& $v$        & 4.388796   & 4.388794   & $< .0001$ & $< .01$\% \\ 
		& $\alpha$   & 13.111988  & 13.111980  & $< .0001$ & $< .01$\% \\ 
		& $m$        & 20.721280  & 20.721270  & $< .0001$ & $< .01$\% \\ 
		\hline
		VI & $\beta_1$  & .995360    &.995361     & $< .0001$ & $< .01$\% \\ 
		& $\beta_2$  & 4.739349   &  4.739349  & $< .0001$ & $< .01$\% \\ 
		& $\kappa$        & 1.894437   & 1.895000   & .0006   & .03\% \\ 
		& $r$        & -33.421430 & -33.421290 & .0001   & $< .01$\% \\  
		& $q_1$      & 42.030520  & 42.030800  & .0003   & $< .01$\% \\ 
		& $q_2$      & 6.609095   & 6.609500   & .0004   & $< .01$\% \\ 
		& $\alpha$   & 10.379832  & 10.379470  & .0004   & $< .01$\% \\ 
		\hline
		\multicolumn{6}{l}{\small{$^a$Elderton and Johnson (1969) does not have the other types of Pearson distributions.}} \\
		\multicolumn{6}{l}{\small{$^b$Absolute Difference = $\lvert$Value from Elderton and Johnson (1969) $-$ Value from \proglang{SAS/IML} Macro$\rvert$.}} \\
		\multicolumn{6}{l}{\small{$^c$Relative Difference = $\lvert$(Value from Elderton and Johnson (1969) $-$ Value from \proglang{SAS/IML} Macro)}} \\
		& & \multicolumn{4}{l}{\small{/Value from Elderton and Johnson (1969)$\rvert\times$100\%.}}
	\end{tabular}
	\caption{Computed parameters and their accuracy.}
	\label{table2}
\end{table}

Then, the computed probability values from the \proglang{SAS/IML} macro were evaluated using the percentage points in \citet{Pearson72}'s Table 32 (p. 276) corresponding to probability values of 2.5\% and 97.5\%. From Table \ref{table3}, we can see that the probability values computed from the \proglang{SAS/IML} macro are very close to .025 (or 2.5\%) and .975 (or 97.5\%), respectively, with a high degree of precision (less than .0001).

\begin{table}[h]
	\centering
	\begin{tabular}{lrrccccccrr}
		\hline
		&&& \multicolumn{2}{c}{Percentage Point} &&&&&& \\
		&&& \multicolumn{2}{c}{from Pearson and} && \multicolumn{2}{c}{Probability Value} && \multicolumn{2}{c}{Absolute} \\
		&&& \multicolumn{2}{c}{Hartley (1972)} && \multicolumn{2}{c}{from \proglang{SAS/IML} Macro} && \multicolumn{2}{c}{Difference$^b$} \\
		\cline{4-5} \cline{7-8} \cline{10-11}
		Type$^a$ & $\surd{\beta_1}$ & $\beta_2$ & For 2.5\% & For 97.5\% && 2.5\% & 97.5\% && \multicolumn{1}{c}{For 2.5\%} & \multicolumn{1}{c}{For 97.5\%} \\
		\hline
		Normal & .0  & 3.0  & -1.9600 & 1.9600 && .0249970 & .9750020 && $< .00001$ & $< .00001$ \\ 
		I      & .6  & 3.2  & -1.5998 & 2.2320 && .0249965 & .9749989 && $< .00001$ & $< .00001$ \\ 
		II     & .0  & 2.6  & -1.9196 & 1.9196 && .0250030 & .9749970 && $< .00001$ & $< .00001$ \\ 
		IV     & 1.4 & 8.6  & -1.5068 & 2.3801 && .0249838 & .9749471 &&   .00002 &   .00005 \\ 
		VI     & 2.0 & 11.2 & -1.1915 & 2.5545 && .0250054 & .9750021 &&   .00001 & $< .00001$ \\ 
		VII    & .0  & 8.4  & -1.9925 & 1.9925 && .0249999 & .9750001 && $< .00001$ & $< .00001$ \\ 
		\hline
		\multicolumn{11}{l}{\small{$^a$Pearson and Hartley (1972) does not have examples of types III and V.}} \\
		\multicolumn{11}{l}{\small{$^b$Absolute Difference = $\lvert$.025 $-$ Probability value from \proglang{SAS/IML} macro$\rvert$, and = $\lvert$.975 $-$ Probability value}} \\
		\multicolumn{11}{l}{\small{ from \proglang{SAS/IML} macro$\rvert$, respectively.}}
	\end{tabular}
	\caption{Computed probability values and their accuracy.}
	\label{table3}
\end{table}

\section{Concluding remarks}
\label{sec:concl}

The new \proglang{SAS/IML} macro program provides an efficient and accurate means to compute probability values of Pearson distributions for which any data can be approximated based on the first four moments of the data. Thus, researchers can utilize this \proglang{SAS/IML} macro program in conducting distribution-free statistical analysis for any data with unknown distributions. The \proglang{SAS/IML} macro program also provides a nice feature of graphing the probability values of Pearson distributions to visualize the probability values on the Pearson distribution curves. For future study, it would be desirable to develop the similar program in other commonly used statistical language such as R or Stata. 

\bigskip
\begin{center}
	{\large\bf Supplementary Material}
\end{center}

\noindent%
The \proglang{SAS/IML} macro program for computing and graphing probability values of Pearson distributions is available as an ancillary file, \textit{PearsonDistributionProb.txt}.

\bibliography{PearsonRef}

\end{document}